\documentclass{eas}
\usepackage{graphicx}
\begin{document}

\title{The population of M-type supergiants in the starburst cluster Stephenson~2} 
\runningtitle{M-type supergiants in Stephenson~2}.
\author{Ignacio Negueruela}\address{Departamento de F\'{\i}sica, Ingenier\'{\i}a de Sistemas y Teor\'{\i}a de la Se\~{n}al, Universidad de Alicante, Apdo. 99, E-03080
  Alicante, Spain}
\author{Carlos Gonz\'alez-Fern\'andez}\sameaddress{1}
\author{Ricardo Dorda}\sameaddress{1}
\author{Amparo Marco}\sameaddress{1}
\author{J. Simon Clark}\address{Department of Physics and Astronomy, The Open 
University, Walton Hall, Milton Keynes, MK7 6AA, UK}
\begin{abstract}
The open cluster Stephenson 2 contains the largest collection of red supergiants known in the Galaxy, and at present is the second most massive young cluster known in the Milky Way. We have obtained multi-epoch, intermediate-resolution spectra around the Ca\,{\sc ii} triplet for more than 30 red supergiants in Stephenson~2 and its surroundings. We find a clear separation between a majority of RSGs having spectral types M0-M2 and the brightest members in the NIR, which have very late spectral types and show strong evidence for heavy mass loss. The distribution of
spectral types is similar to that of RSGs in other clusters, such as NGC~7419, or associations, like Per~OB1. The cluster data strongly support the idea that heavy mass loss and maser emission is preferentially
associated with late-M spectral types, suggesting that they represent an evolutionary phase.
\end{abstract}
\maketitle
\section{Introduction}

Red supergiants (RSGs) are very difficult to study in isolation. Accurate stellar parameters depend on a good estimation of their distances, while comparison of these parameters to evolutionary models requires knowledge of their ages. Such information is not available for isolated RSGs.

Observing extragalactic populations (e.g., \cite{levmassey12}) removes the distance problem and provides large, statistically-significant samples, but may introduce other complications  (fainter targets, spatially unresolved systems, contamination by bright AGB stars, possible metallicity gradients, mixed ages, ...). 

For Galactic stars, distances and ages are inferred from membership in open clusters or OB associations (\cite{levesque}). 
Unfortunately, given the short duration of the RSG phase, typical open clusters in the solar neighbourhood, with total initial masses $\sim10^{3}\:M_{\odot}$, do not contain more than one or two RSGs. Indeed, simulations of open clusters using modern stellar evolution models and a standard IMF suggest that cluster masses $\sim10^{4}\:M_{\odot}$ are needed to have a good statistical chance of finding 2\,--\,4 RSGs (\cite{clark09a}). Therefore population analysis must be done by combining large cluster samples (e.g., \cite{eggenberger}; \cite{levesque}) and thus mixing RSGs of different ages and perhaps even somewhat different chemical compositions. 

Mixed populations can be used to derive the current properties of the RSGs, but provide only weak constraints on their parent populations. Stronger constraints on evolutionary models could be obtained by analysing moderately-sized single-age populations of RSGs. The Perseus Arm contains some interesting examples of such populations (see \cite{humphreys70}). The Per~OB1 association includes almost 20 RSGs, but its exact boundaries are not certain -- not only its spatial extent, but also its actual membership (cf. \cite{walborn02}). The core region, centred on the double cluster $h$ and $\chi$~Persei, contains $\sim10$ RSGs for a total mass $\sim 20\,000\:M_{\odot}$, though again measuring the total mass over an area $\sim1^{\circ}$ across is a difficult process (\cite{currie10}).
 
The best example of a well-defined RSG population is provided by NGC~7419. This compact open cluster, with a mass $\geq5\times10^{3}\:M_{\odot}$ contains 5 RSGs at an age $14\pm2$~Myr (\cite{marco}). One of its most remarkable features is the presence of the very late supergiant MY~Cep. This object, which presents strong evidence for heavy mass loss (\cite{fawley};\cite{verheyen}), seems to have been spectroscopically stable at spectral type M7.5\,I for the last $\sim50$~yr. Such a late spectral type is very unusual for an RSG. Moreover, the few other RSGs observed with very late ($>$M4) spectral types tend to spend most of the time at earlier spectral types, and go to late-M types only during brief excursions (e.g., VX~Sgr, which displays excursions up to M10; see \cite{wing09}).

In an analysis of the Milky Way RSG population, Levesque {\em et al}. (2005; see also  Levesque, in these proceedings) found M-type supergiants with spectral types between M0 and M5, with a clear peak around spectral type M2. This result is based on a sample of RSGs in OB associations. There is some hint that the RSGs in open clusters have a slightly different distribution of spectral types, preferentially having spectral types in the M0-M2 range, with a small fraction of them having later spectral types and indications of heavy mass loss. For example, in NGC~3766, the two RSG members have spectral type M0\,Ib (\cite{humphreys78}). In $\chi$~Per, FZ~Per, BD~$+56^{\circ}$595 and HD~14580 are M0-1\,Iab, AD~Per is M2.5\,Iab and RS~Per is M4\,Iab (\cite{humphreys70}). In NGC~7419, four RSGs are M0-2\,Iab and MY~Cep is much later. Unfortunately, the number of open clusters with more than one RSG is small, and all these clusters contain relatively small populations of RSGs. Once again, coming to any conclusion implies combining data from many different clusters and ignoring potentially complicating factors, such as an age dependence.

\section{Stephenson~2}

The advent of infrared detectors has allowed the exploration of the inner Milky Way, leading to the discovery of many sites of intense star formation and many massive young star clusters. In the last few years, several clusters containing many RSGs have been found towards the base of the Scutum Arm (\cite{figer06}; \cite{clark09b}). The open cluster Stephenson~2, though already known and identified as a candidate massive star cluster (\cite{stephenson}), turned out to contain the highest number of RSGs (\cite{davies07}). This cluster lies behind a dust layer (LDN 515) and is thus affected by heavy extinction ($A_{V}\approx11$~mag).

Davies {\em et al}. (2007) obtained $K$-band spectra of sources within a circle of radius $7^{\prime}$, finding a large number of RSGs with similar radial velocities. By assuming that stars with radial velocities within $\pm10\:{\rm km}\,{\rm s}^{-1}$ of the average value were cluster members, they concluded that 26 RSGs belonged to the cluster. The average radial velocity $v_{{\rm LSR}}=+109.3\pm0.7\:{\rm km}\,{\rm s}^{-1}$ implies a kinematic distance $\approx 6$~kpc. If the cluster is at that distance, its age is $\sim17$~Myr, and its total mass must be $>5\times10^{4}\:M_{\odot}$. From the depth of the CO bandhead at $2.3$~$\mu$m, Davies {\em et al}. (2007) estimated luminosity and spectral type for all their stars, as this feature is sensitive to both parameters.

The reddening to Stephenson~2, though high, still allows observation of its RSGs in the region around the Ca\,{\sc ii} triplet. This spectral region provides much more information than the $K$ band. Classification criteria for RSGs have been known since the photographic plate era (\cite{kh45}; \cite{sharpless}) and have been extended to digital spectra (\cite{carquillat97}), though most of them are only valid for spectral types up to M3. Later types are not so well characterised, but criteria have been defined (e.g., \cite{solf}; see also Dorda {\em et al}.\ in these proceedings). Observing a wide area around the cluster with the multi-object fibre spectrograph AutoFib2+WYFFOS (AF2) mounted on the Prime Focus of the 4.2~m William Herschel Telescope (WHT; La Palma, Spain), Negueruela {\em et al}. (2012) found many more RSGs with radial velocities similar to the cluster average. They concluded that the spatial extent of Stephenson~2 cannot be clearly defined. A compact core, with radius $\leq2.5^{\prime}$ and containing about 20 RSGs, is surrounded by an extended association that merges into a general background with an important over-density of RSGs. 

\section{Observations}

We have complemented the observations reported in Negueruela {\em et al}. (2012) with new observations at higher resolution. Spectra of most RSGs in Stephenson~2 were obtained during three runs in 2010, 2011 and 2012 with the fibre-fed dual-beam AAOmega spectrograph mounted on the 3.9~m Anglo-Australian Telescope (AAT) at the Australian Astronomical Observatory. The instrument was operated with the Two Degree Field ("2dF") multi-object system as front-end.
We used the red arm, suited with a low-fringing CCD, equipped with the 1700D grating, blazed at 10\,000\AA, and specifically designed to observe the Ca\,{\sc ii} triplet region. This grating provides a resolving power $R=10\,000$ over slightly more than 400\AA. The exact wavelength range observed depends on the position of the target in the 2dF field.

Further observations were obtained in June 2012 with the red arm of the long-slit ISIS double-beam spectrograph, also mounted on the WHT. We used grating R1200R centred on 8700\AA. The instrument was fitted with the {\it Red+} CCD, which has very low fringing in this range and covered the
8400\,--\,8950\AA\ range with a nominal dispersion of 0.26\AA/pixel. We used a $0.9^{\prime\prime}$ slit, resulting in a resolving power $R\approx16\,000$.

\section{Red supergiants in Stephenson~2}

\begin{figure}
  \centering
\resizebox{\columnwidth}{!}{\includegraphics[clip]{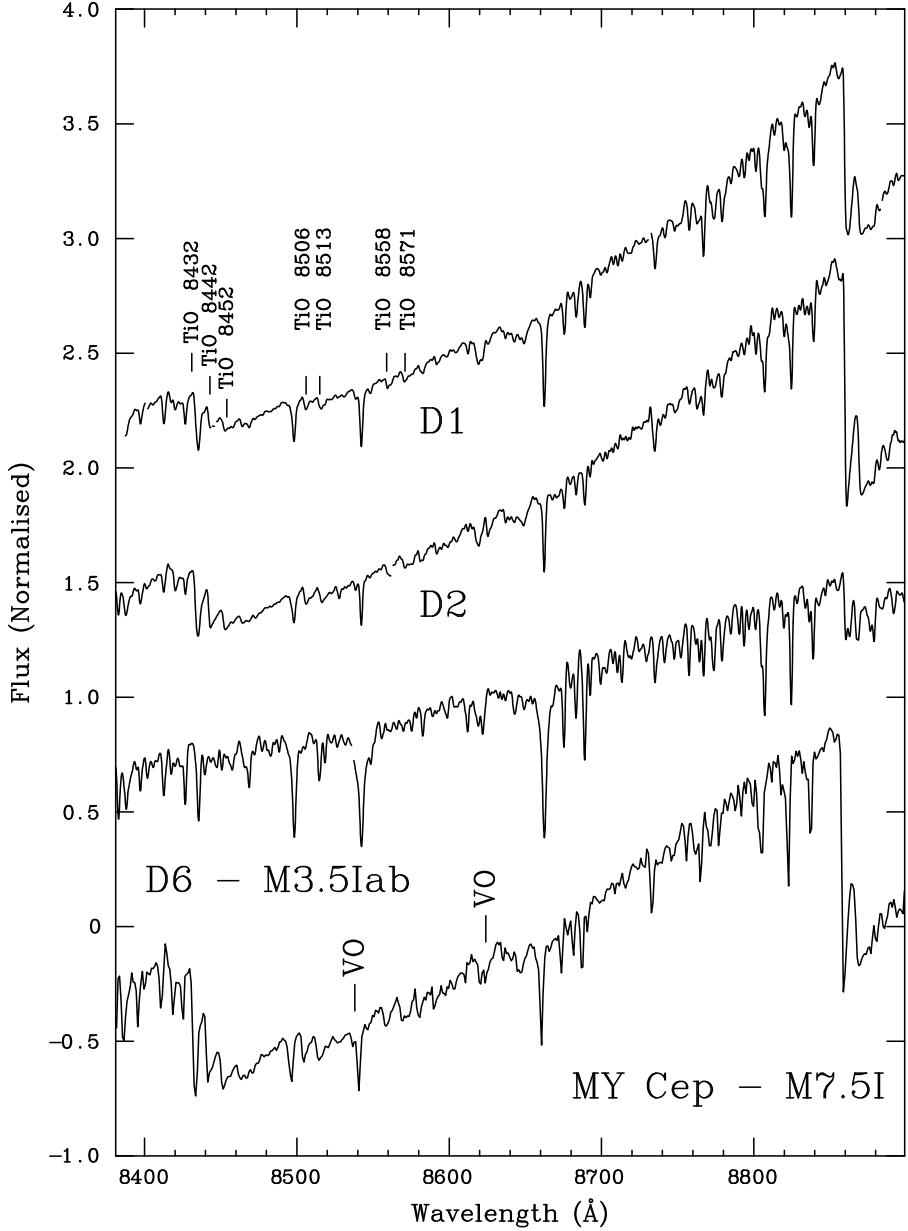}}
   \caption{Three RSGs in Stephenson~2 compared with the very late supergiant MY~Cep, a member of the open cluster NGC~7419. The shape of the spectra for stars in Stephenson~2 is determined by the very strong reddening ($A_{V}\sim11$~mag). The presence of VO bandheads at 8538, 8624\AA\ (shown on the spectrum of MY~Cep) places D2 at a spectral type not earlier than M7. Given the smaller TiO bandhead at 8860\AA, D1 is slightly earlier (around M6). Other prominent TiO bandheads are marked on its spectrum. \label{fig}}
    \end{figure}

Of the 28 RSGs listed by Davies {\em et al}. (2007), we have observed 24. One supergiant (D49) is very heavily reddened, and is not detectable in the $I$ band down to the DENIS limiting magnitude ($i\approx18$~mag). The two faintest RSG members (D52 and D72) are not detected in DENIS either (perhaps due to crowding). Finally, D23, which is an outlier in terms of spatial location and radial velocity, is too faint for a spectrum in this spectral region.

Our data fully confirm the nature of all the other supergiants, validating the results obtained by Davies {\em et al}. (2007) with the CO bandhead technique. The spectral types derived from the $K$-band spectra, however, are only approximate. The CO-bandhead calibration results in a small average dispersion, but a given star may easily be wrong by 3 subtypes (\cite{negueruela12}).

Two of the brightest supergiants, D1 and D4, have CO-bandhead radial velocities not consistent with cluster membership, but not far from the cluster average (\cite{davies07}). Our radial velocities, however, indicate membership for D1, the brightest star in the field (in the $K$ band). The radial velocity of D4 is still outside the range adopted for cluster members, but there are some reasons to think that this star belongs to the cluster (see below).

Assuming that the very high extinction to D49 is interstellar, Davies {\em et al}. (2007) calculated that this is the intrinsically brightest member of the cluster, slightly more luminous than D2. If D1 is indeed a member, it is much brighter than any of the two. With $(J-K_{{\rm s}})\approx7~mag$, D49 is a very bright mid-IR source (unfortunately, it seems to be saturated in both the {\it GLIMPSE} and {\it WISE} catalogues). It is thus very likely that this RSG is embedded in a thick envelope and actually self-absorbed.  It displays SiO and H$_{\rm 2}$O maser emission, generally considered strong evidence of mass loss into an extended envelope (\cite{deguchi}; \cite{verheyen}).

Maser emission has also been detected from D1 and D2, but not from any other member (\cite{deguchi}; \cite{verheyen}), suggesting that these objects have the highest mass loss rates amongst RSGs in Stephenson~2. Only the most luminous cluster members (in terms of bolometric luminosity) seem to present detectable maser emission.

Our spectra (see Fig.~1) show that D2 has the latest spectral class in the cluster. It may be slightly variable, oscillating around M7\,I. D1 is very slightly earlier. D3, D4 and D5 have spectral type M5\,I, while D6 and D9 are M3.5\,Iab, showing a remarkable correlation between brightness and spectral type (note that the numbering system of \cite{davies07} reflects brightness in $K_{{\rm s}}$). The only exception is D8, which might not be a real member, as it does look like a very late-type giant. All stars fainter than D11 for which we have spectra (14 objects), have spectral types between M0\,I and M2.5\,I, with most concentrating around M1.5\,Iab. 

The distribution of spectral types in Stephenson~2 is thus consistent with the information from other clusters, suggesting that the M\,I subtypes represent, to a first approximation, some sort of evolutionary sequence. Given the observed fractions, most stars should spend a substantial fraction of their time as RSGs at spectral type M1--M2 and then move to later spectral types, becoming brighter. Extreme mass loss, as reflected by maser emission or very bright mid-IR colours, seems to happen only at later spectral types. RSGs in the Per~OB1 association also show a similar correlation between spectral type and luminosity and between spectral type and mass loss, measured from the mid-infrared excess (\cite{cohen}). Field RSGs in the Carina arm display similar correlations (\cite{humphreys74}).

This view would be consistent with recent results suggesting that all M-type RSGs have similar effective temperatures and that the strength of the TiO spectrum, the main criterion used for spectral classification, is strongly correlated to luminosity rather than temperature (\cite{davies13}; and also in these proceedings). The evolution towards later spectral types would then be caused by the heavier mass loss, which would result in the production of more extended molecular layers as the star's luminosity increases.

The clusters of RSGs will be excellent laboratories to test these hypotheses, as they provide large populations of RSGs. In a given cluster, all RSGs are expected to have the same age and thus similar initial masses (though the effects of high initial rotation may lead to some mixing of stars with different initial masses; see \cite{ekstrom}). The comparison between different clusters can help disentangle any age dependence of the average spectral types. For this, accurate distances and ages to the cluster still have to be determined. 

\section*{Acknowledgments}
This research is partially supported by the Spanish Ministerio de Econom\'{\i}a y Competitividad (Mineco) under grant AYA2010-21697-C05-05, and by the Generalitat Valenciana (ACOMP/2012/134).  The WHT is operated on the island of La
Palma by the Isaac Newton Group in the Spanish Observatorio del Roque
de Los Muchachos of the Instituto de Astrof\'{\i}sica de Canarias. This research has made use of the Simbad database, operated at CDS,
Strasbourg (France) and of the WEBDA database, operated at the
Institute for Astronomy of the University of Vienna. 


\end{document}